\documentclass[
 reprint,
superscriptaddress,
amsmath,amssymb,
aps,
]{revtex4-2}

\usepackage{graphicx}
\usepackage{dcolumn}
\usepackage{bm}
\begin{document}

\title{
An integrated atom array - nanophotonic chip platform with background-free imaging
}

\author{Shankar G. Menon}
\thanks{These authors contributed equally to this work}
\affiliation{
 Pritzker School of Molecular Engineering, University of Chicago, Chicago, IL 60637, USA
}
\author{Noah Glachman}
\thanks{These authors contributed equally to this work}
\affiliation{
 Pritzker School of Molecular Engineering, University of Chicago, Chicago, IL 60637, USA
}
\author{Matteo Pompili}
\affiliation{
 Pritzker School of Molecular Engineering, University of Chicago, Chicago, IL 60637, USA
}
\author{Alan Dibos}
\affiliation{
Center for Nanoscale Materials, Argonne National Laboratory, Lemont, IL 60439
}
\affiliation{Nanoscience and Technology Division, Argonne National Laboratory, Lemont, IL 60439}
\affiliation{Center for Molecular Engineering, Argonne National Laboratory, Lemont, IL 60439}
\author{Hannes Bernien}
\email{bernien@uchicago.edu}
\affiliation{
 Pritzker School of Molecular Engineering, University of Chicago, Chicago, IL 60637, USA
}

\begin{abstract}

Arrays of neutral atoms trapped in optical tweezers have emerged as a leading platform for quantum information processing and quantum simulation due to their scalability, reconfigurable connectivity, and high-fidelity operations. Individual atoms are promising candidates for quantum networking due to their capability to emit indistinguishable photons that are entangled with their internal atomic states. Integrating atom arrays with photonic interfaces would enable distributed architectures in which nodes hosting many processing qubits could be efficiently linked together via the distribution of remote entanglement. However, many atom array techniques cease to work in close proximity to photonic interfaces, with atom detection via standard fluorescence imaging presenting a major challenge due to scattering from nearby photonic devices. Here, we demonstrate an architecture that combines atom arrays with up to 64 optical tweezers and a millimeter-scale photonic chip hosting more than 100 nanophotonic devices. We achieve high-fidelity ($\sim$99.2\%), background-free imaging in close proximity to nanofabricated devices using a multichromatic excitation and detection scheme. The atoms can be imaged while trapped a few hundred nanometers above the dielectric surface, which we verify using Stark shift measurements of the modified trapping potential. Finally, we rearrange atoms into defect-free arrays and load them simultaneously onto the same or multiple devices.
\end{abstract}

\maketitle

\begin{figure*}
\includegraphics{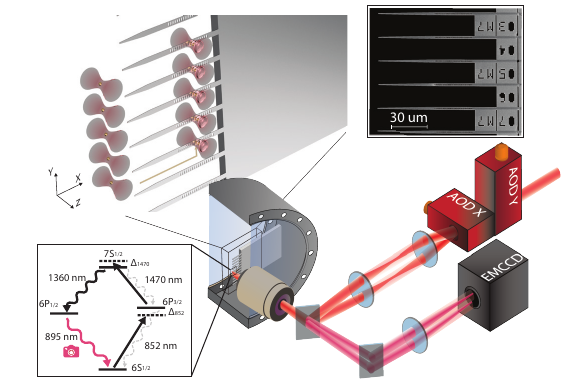}
\caption{\label{fig1} \textbf{The atom array - nanophotonic chip platform.} Schematic of the platform depicting our optical tweezer array manipulating single cesium atoms near a chip hosting an array of nanophotonic devices. The zoomed-out view shows the chip inside a stainless steel ultra-high vacuum chamber along with the key optical components used to control and image the system (AOD: acousto-optic deflector, EMCCD: electron multiplying charge-coupled device). The top inset shows a scanning electron microscope (SEM) image of the nanophotonic devices at the edge of our chip. The bottom inset shows the fine level structure that illustrates our background-free imaging technique. We doubly excite the atoms from the 6S$_{1/2}$ ground state following the straight black arrows, through the 6P$_{3/2}$ intermediate state, and spectrally filter the fluorescence reaching the camera to only image the 895 nm decay path from the 6P$_{1/2}$ level shown in red. Here, $\Delta_{852}$ is the detuning from the bare 6S$_{1/2}$ F=4 $\rightarrow$ 6P$_{3/2}$ F$'$=5 transition and $\Delta_{1470}$ is the detuning from the bare 6P$_{3/2}$ F$'$=5 $\rightarrow$ 7S$_{1/2}$ F$''$=4 transition.}
\end{figure*}

Neutral atom arrays trapped in optical tweezers show promise as quantum information processors due to their characteristic scalability \cite{Ebadi2020, Scholl2021} and programmability with any-to-any connectivity \cite{Bluvstein2021, Graham2022}. Recently, high-fidelity entangling operations \cite{Evered2023, Scholl2023, Ma2023} and the capability to perform mid-circuit readout and feedback \cite{Singh2023, Graham2023, Lis2023, Norcia2023} have also been demonstrated in atom-array systems. Combining these systems with a photonic interface can further enhance their capabilities by enabling quantum communication \cite{kimble2008}, blind and distributed quantum computation \cite{broadbent2008,Arrighi2006,Drmota2023,Monroe2014}, mid-circuit and faster readout techniques \cite{Deist2022}, distributed sensing \cite{Komar2014, Gottesman2012}, as well as by extending the set of long-range interaction Hamiltonians that can be simulated \cite{Douglas2015, Bello2019, Tudela2015, Mahmoodian2020, Chang2013, Fayard2021, Yan2023}.

Trapped atoms have been integrated with mirror cavities \cite{Mabuchi1998, Bochmann2008,Purdy2010,Liu2023}, fiber cavities \cite{Brekenfeld2020}, nanofiber-based waveguides \cite{ Vetsch2010}, as well as nanophotonic waveguides and cavities \cite{Goban2014,Thompson2013}. Among these, nanophotonic cavities and waveguides offer some of the highest atom-photon interaction strengths owing to their small mode volumes and high quality factors \cite{Samutpraphoot2020}. Up to two atoms have been deterministically trapped on top of a nanophotonic cavity and entangled using cavity-carving techniques \cite{Samutpraphoot2020, Dordevic2021}.

Scaling systems that combine atoms with photonic interfaces and incorporating atom array capabilities is not a straightforward task. The presence of a dielectric structure in the vicinity of the atomic traps can significantly impact the atom loading probability, while undesired scattering from the nanophotonics makes detecting the atoms via standard fluorescence imaging extremely difficult. Several approaches have been used to image atoms in the vicinity of such structures including confocal microscopy \cite{Samutpraphoot2020}, exciting atoms through propagating modes of the waveguides \cite{Meng2020}, and polarization filtering combined with spatial filtering \cite{Kim2019}, however, these techniques have been limited to either few atoms or specific device geometries that are not broadly applicable. Thus, demonstrating atom array techniques such as rearrangement and single-shot readout near arbitrary nanophotonic devices remains an outstanding challenge. The presence of nearby dielectric structures can also have strong effects on the highly excited Rydberg states of the atoms, degrading the performance of Rydberg mediated gates near dielectric interfaces. However, recent work has shown that reliable Rydberg gates can be performed in the vicinity of a dielectric surface by moving atoms tens of microns away \cite{Ocola2022}.

In this work, we introduce an experimental platform combining an atom array with a silicon nitride on silicon chip hosting more than 100 nanophotonic devices along with a background-free imaging scheme to overcome scattering from the nearby chip. A multichromatic imaging technique \cite{background_free_linke_2012, background_free_mcgilligan_2020, background_free_ohadi_2009, background_free_yang_2012} suppresses the device scattering, enabling single-shot readout of the entire atom array close to, or even on top of, the devices using an electron-multiplying charge-coupled device (EMCCD) camera similar to the standard readout method for free-space atom arrays. We find that the presence of our chip has minimal effects on the atom loading characteristics and demonstrate that we can rearrange atoms and load them onto multiple devices at the same time or to a single device, where they can be imaged in a single shot with our imaging technique. 

This platform combines the measurement and rearrangement capabilities of atom arrays with the ability to engineer the photonic environment via integrated cavities and waveguides, representing an enabling step towards multiplexed telecom quantum networking with resonant cavities \cite{Menon2020, Huie2021}, fault-tolerant distributed quantum computing with Rydberg integration \cite{Ramette2023}, and demonstrations of novel many-body phenomena in atom-waveguide systems including self-organization of atoms and the generation of arbitrary photonic states \cite{Chang2013, Tudela2015}.

\section*{Results}

\begin{figure*}
\includegraphics{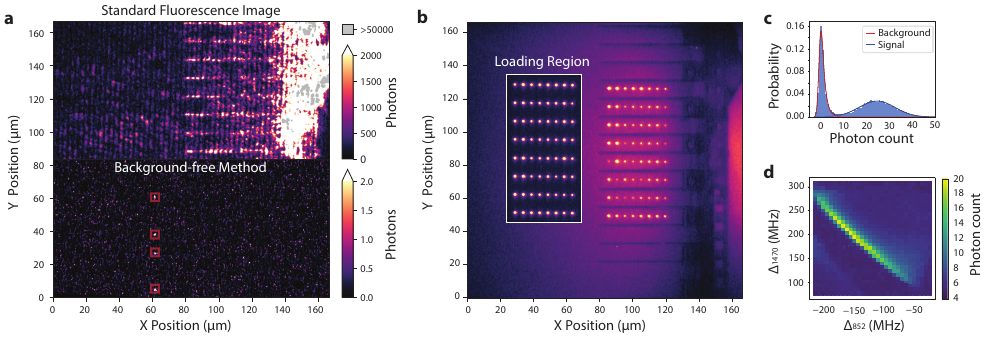}
\caption{\label{fig2} \textbf{Background-free imaging of the atom array.} \textbf{a}, Top: Fluorescence image using the standard 852 nm (D2) cycling transition near our nanophotonic chip. Despite lowering the electron multiplying gain (EM gain) process on our camera to 10 from our typical value of 1000, the image is still saturated at the single atom scale even tens of microns away from the devices. Bottom: A single-shot image of atoms (inside the red boxes) taken using the background-free technique developed in this work. Here, both images are taken with a 40 ms exposure time.  \textbf{b}, An averaged fluorescence image of the atom array interleaved between the nanophotonic device array. No post-processing has been applied here beyond averaging the individual images. A small residual background makes the devices appear as dark shadows. We attribute this background to fluorescence at our imaging wavelength from the silicon base layer of the nanophotonic chip. The inset shows an averaged fluorescence image of an 8x8 atom array in the loading region. \textbf{c}, A typical histogram of the detected 895 nm photons within a 4$\times$4 pixel region of interest over a 40 ms exposure time in the loading region. The bimodal distribution allows us to distinguish between the presence and absence of an atom with high fidelity $\gtrsim$99.2\%. \textbf{d}, Average magnitude of the detected fluorescence signal from the atoms as a function of the two drive laser detunings. The signal falls off as the 852 nm laser approaches hyperfine resonances on each side of the plot, leading to atom loss from the tweezers.}
\end{figure*}

Our platform consists of an array of optical tweezers formed by a pair of crossed acousto-optic deflectors (AODs) that can be used to trap, move, and rearrange individual atoms in our ultra-high vacuum chamber as depicted in Fig. \ref{fig1}. We load our tweezers with cesium atoms $\lesssim$100 $\mu$m away from a 600 $\mu$m thick 2 mm $\times$ 8 mm chip hosting over one hundred nanophotonic crystal cavities $\sim$60 $\mu$m long (Fig. \ref{fig1} inset), though the techniques and results presented here are independent of the device type and broadly apply to any dielectric structures. The devices are fabricated from 330 nm thick silicon nitride on top of a silicon substrate, which is completely undercut from the device region, enabling the tweezer beams to pass by the chip even in close proximity to the devices (for details see supplementary information). We cool the atoms by forming a magneto-optical trap (MOT) and load atoms into the optical tweezers from the MOT in a loading region tens of microns from the devices (see Fig. \ref{fig2}\textbf{b}). The optical beams that form the MOT and image the atoms are partially reflected by the chip and devices, rendering the standard fluorescence imaging technique of driving the 852 nm cycling (D2) transition impractical as the scattering from the surface of the solid structure is many orders of magnitude more intense than the atomic fluorescence. We demonstrate the scale of this problem in Fig. \ref{fig2}\textbf{a}, where we show that even away from the devices, in our loading region, this scattering leads to signals on the order of hundreds to thousands of photons per pixel, compared to our atomic signals which are typically on the order of $\sim$1.6 photons per pixel (25 photons detected within a 4$\times$4 pixel region of interest). On the region of the devices, we have several pixels detecting more than 350,000 photons. At higher EM gain settings necessary for imaging single atoms, signals of this magnitude would rapidly damage the camera sensor.

We use a background-free imaging technique to circumvent this scattering issue where we drive a two-photon transition to an excited state (here the 7S$_{1/2}$ state), which has two decay pathways (see Fig. \ref{fig1} bottom inset). We spectrally filter out the excitation wavelengths and image the fluorescence from the atoms along the other decay path at 895 nm. With this method, we filter out all of the unwanted background due to the scattering from the devices and achieve high detection fidelities (see Fig. \ref{fig2}\textbf{c}). We show an averaged fluorescence image of our array both away from the devices in the loading region (Fig. \ref{fig2}\textbf{b} inset) and interleaved between the devices (Fig. \ref{fig2}\textbf{b}). A typical histogram of the signal received from a single site (from Fig. \ref{fig2}\textbf{b} inset) showing the ability to discriminate with high fidelity between the presence and absence of an atom is shown in Fig. \ref{fig2}\textbf{c}. We estimate an imaging fidelity of 99.2\% by fitting the bimodal histogram and examining the overlap of the signal and the background. Here, we use a 40 ms exposure time, comparable to the timescale of standard fluorescence imaging \cite{Endres2016}, making this method applicable to a wide range of experimental setups that suffer from background light. 

In Fig. \ref{fig2}\textbf{d}, we plot the magnitude of the fluorescence signal detected from the atoms as we vary the detunings of the two drive lasers. The 852 nm laser detuning on the x-axis is measured with respect to the bare 6S$_{1/2}$ F=4 $\rightarrow$ 6P$_{3/2}$ F$'$=5 transition and the 1470 nm laser detuning on the y-axis is measured with respect to the bare 6P$_{3/2}$ F$'$=5 $\rightarrow$ 7S$_{1/2}$ F$''$=4 transition. The detected signal falls off as the 852 nm detuning approaches the 6S$_{1/2}$ F=4 $\rightarrow$ 6P$_{3/2}$ F$'$=5 transition on the right side of the plot and when it approaches the 6S$_{1/2}$ F=4 $\rightarrow$ 6P$_{3/2}$ F$'$=4 transition (-251 MHz on the x-axis) on the left side of the plot. We attribute this to atom loss from the tweezer due to resonant heating. The optimal point is far from a ground state resonance, where cycling on the lower transition is suppressed and the two-photon excitation to the doubly excited state is the dominant process. The overall feature is blue-shifted from the bare two-photon resonance by tens of MHz due to the AC Stark shift from the tweezer.

\begin{figure*}
\includegraphics{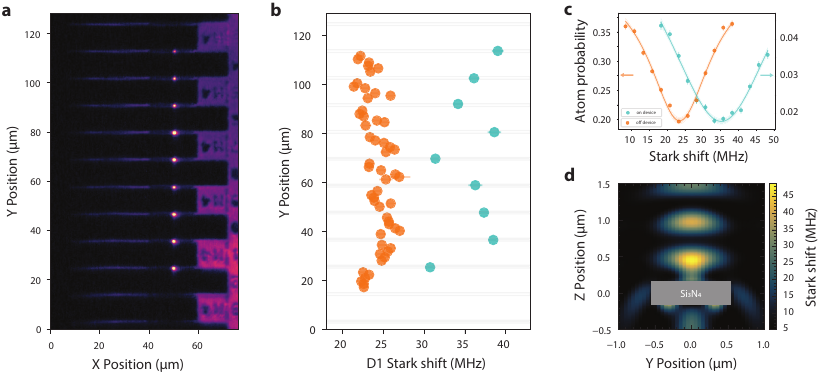}
\caption{\label{fig3} \textbf{Imaging atoms on top of devices.} \textbf{a}, The averaged fluorescence from 15,000 individual images of atoms loaded onto the nanophotonic devices overlaid with an image of the devices (see methods for image processing details). \textbf{b}, The centers of the Stark shift curves for individual atoms as a function of their positions. The grey lines are estimated device positions from Fig. \ref{fig3}\textbf{a}. The Stark shifts are larger in magnitude when the atoms are on top of the devices as expected from the modified trapping potential on top of the devices as shown in \textbf{d} (for the trapping potential away from the devices, see supplementary information). \textbf{c}, Averaged Stark shift measurements on the 895 nm 6S$_{1/2}$ F=3 $\rightarrow$ 6P$_{1/2}$ F$'$=4 transition in between the devices (orange) and on top of the devices (cyan). Centers estimated from Lorentzian fits to similar plots at each position are used to generate \textbf{b}. \textbf{d}, Cross section of the nanophotonic device and the expected Stark shift on the D1 transition for the different intensity maxima formed on top of the device by the reflected tweezer.}
\end{figure*}

Another crucial capability required for nanophotonic integration is the ability to move atoms onto the photonic chip in close enough proximity to couple the atoms to the light field of the devices. We achieve this by adiabatically translating the optical tweezers that are loaded with individual atoms from the loading region until they aim directly at the devices, where the tweezer beams partially reflect backward, forming standing wave traps above the devices as shown in Fig. \ref{fig3}\textbf{d}. We show an averaged image of the atoms loaded on top of the devices overlaid with an image of the devices (see methods for image processing details) in Fig. \ref{fig3}\textbf{a}. The goal is to load the atoms into the closest intensity maximum $\sim$300 nm from the surface of the silicon nitride, where they can strongly couple to the cavity mode. The distance between the closest intensity maximum and the surface is set by a combination of the tweezer wavelength and the thickness of the device. This standing wave trap has an intensity $\sim$2× the intensity of the tweezer in free space. In order to determine whether we are loading the atom to the standing wave trap, we probe the AC Stark shift experienced by the atom, which is proportional to the intensity of the trap, on the 6S$_{1/2}$ F=3 $\rightarrow$ 6P$_{1/2}$ F$'$=4 (D1) transition at 895 nm and compare it to the Stark shift the atom experiences in the free-space tweezer. We probe this specific transition because our tweezer wavelength of $\lambda$=935 nm is magic for the 6S$_{1/2}$ $\rightarrow$ 6P$_{3/2}$ transition, rendering it insensitive to the intensity variation between the trapping potentials.

We begin this Stark shift measurement by loading the tweezers with atoms in the loading region, moving the tweezers between the devices, and then moving the tweezers onto the devices from the side. We then apply a variable frequency 895 nm laser pulse in order to blow out the atoms from the tweezer when the pulse is resonant with the Stark-shifted atomic transition. Finally, we image the atoms to detect the survival rate. In Fig. \ref{fig3}\textbf{c}, we show typical blow-out survival curves showing the increased Stark shift when the atoms are loaded onto the devices. In Fig. \ref{fig3}\textbf{b}, we plot the fitted centers of the blow-out survival curves as we move the atoms across the device region, showing that the observed increase in the Stark shift only occurs when atoms are directly on top of the devices where they are trapped in the higher intensity standing wave potential shown in Fig. \ref{fig3}\textbf{d}. From the observed Stark shifts and blow-out curves of the individual atoms, our modeling indicates that we load the atoms into the desired first intensity maximum up to 29\% of the time (see supplementary information for details). The tweezer power, aberrations, and the angle between the devices and the tweezer focal plane contribute to the variations in the observed Stark shift and loading probability across the atoms and can be further optimized in future experiments.

\begin{figure*}
\includegraphics{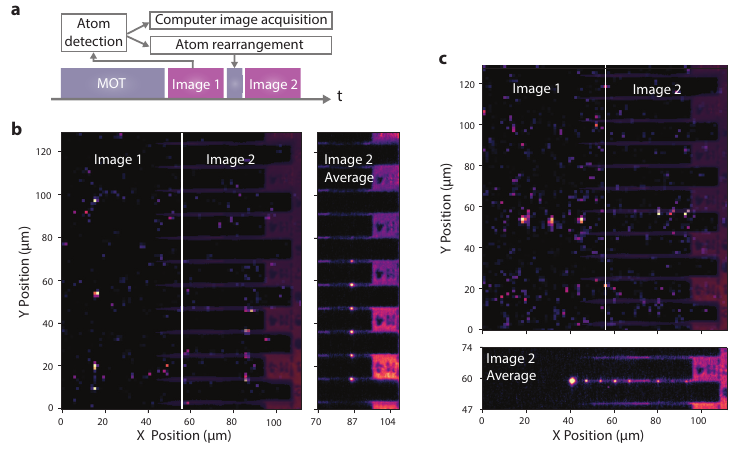}
\caption{\label{fig4}\textbf{Defect-free atomic rearrangement and loading onto devices.}  \textbf{a}, Experimental sequence used for defect-free rearrangement of the atoms from free-space onto the devices. After cooling the atoms and loading them into the tweezers, we take Image 1 to detect the stochastic loading pattern and use this information to rearrange the array into a defect-free configuration which we then translate over and onto the devices before we take Image 2 (see methods for details). \textbf{b}, Image 1 shows a single-shot image of the randomly loaded atoms in a nine tweezer array in the loading region. After detecting these atoms in the first image, they are rearranged into a defect-free array with the same spacing as the devices. This compressed array is then loaded onto the devices and the final configuration is shown in Image 2. An averaged image of the final configuration after this procedure is shown to the right (see methods for image processing details). For all images in this figure, a 25 ms exposure time was used to increase the atomic survival rate through the first image. \textbf{c}, Here, an array of stochastically loaded atoms in Image 1 are rearranged and loaded onto a single device as shown in Image 2. In the bottom plot, the averaged image of the final configuration is shown. The bright single atom on the left of the averaged image is an atom that is rearranged to a position outside the device region.}
\end{figure*}

In order to fully integrate atom arrays with nanophotonics, we must be able to rearrange the atoms into defect-free arrays and then load them onto the devices. To achieve this, we start by taking an image of the initial random loading (25 ms) and rapidly process ($\sim$7 $\mu$s) the image data into an occupation matrix. This information is then used to drop the unoccupied tweezers and compresses the remaining atoms into a defect-free array, before translating the whole array to the devices (see methods for details). The experimental sequence for this procedure is shown in Fig. \ref{fig4}\textbf{a}.

The final step of this procedure can be done either by loading one atom per device or by loading multiple atoms to a single device. We demonstrate both capabilities in Fig. \ref{fig4} by showing initial images of the randomly loaded atoms in free-space tweezers next to a second image of the same atoms post-rearrangement loaded onto the devices in each configuration, one atom per device (Fig. \ref{fig4}\textbf{b}) and three atoms on a single device (Fig. \ref{fig4}\textbf{c}). We also show averaged post-rearrangement images for 15,000 stochastic loading and rearranging events in each configuration, where seven to eight atoms can be seen in each of the post-rearrangement averaged images from the nine initial loading sites.

\section*{Discussion}

We have realized a platform for combining the capabilities of atom arrays with an integrated chip capable of hosting many nanophotonic devices. We load large arrays of atoms near the nanophotonic chip and image them with fidelities greater than 99.2\% with our background-free imaging technique. This capability enables us to load and image atoms on top of the nanophotonic cavities and extract information about the Stark shift to demonstrate atom loading onto the first few intensity maxima on top of the devices. With further optimizations of atomic temperature, tweezer parameters during loading, and device parameters, atoms can be deterministically loaded to the intensity maximum of interest \cite{Thompson2013}. The rate of deterministic placement of atoms in Fig. \ref{fig4} is currently limited by atom survival through the first image, probability of successfully loading to the device, and survival during imaging on top of the devices. A small angle between the device plane and the tweezer plane also causes variation in loading probability across the devices. Improved device alignment and techniques such as stroboscopic imaging or Raman cooling of the atoms can be used to further improve these metrics. 

The devices presented in this work are nanophotonic cavities embedded in waveguides. In future work, these cavities can be utilized to entangle a subsection of the atoms with photons enabling multiplexed entanglement generation \cite{Menon2020,Huie2021}. Furthermore, with the inherent rearrangement capability in our system, atoms can be moved sufficiently far away from the device for Rydberg-mediated gates \cite{Ocola2022}. Incorporating these capabilities would readily enable quantum simulation and computation operations while preserving the ability to distribute remote entanglement, providing a path towards multiplexed quantum repeaters and fault-tolerant distributed quantum computation \cite{Ramette2023}. 

Furthermore, the chip architecture used in this work allows for the simultaneous incorporation of various nanostructures for different purposes hosted on the same chip. Waveguides for quantum simulation and cavities for atom-photon entanglement can be incorporated into the same chip for distributed quantum simulation. Currently, 99\% of the chip surface is not utilized, enabling further opportunities such as the integration of beam splitters, modulators, and detectors directly on the chip.

\section*{Methods}
\subsection*{Platform Details}

Our experiment starts with $\sim$170 ms of MOT loading to maintain a 1:1.5 MOT on/MOT off ratio in our experiment cycles (see supplementary information). The atoms are sourced from a heated dispenser in the same chamber located $\sim$5 cm from the MOT position. This is followed by 10 ms of polarization gradient cooling (PGC) during which atoms are stochastically loaded into the tweezers. We observe 55\% loading efficiency into the tweezer array and atomic temperatures of approximately 50 $\mu$K. A home-built ECDL laser at 935 nm and amplified using a tapered amplifier (MOGLabs MOA) is used to form the optical tweezer array. The RF frequencies that drive the AODs (AA Opto-Electronic) are generated using an AWG for 8$\times$8 array of tweezers and an FPGA for 1$\times$9 and 9$\times$1 arrays (Quantum Machines, OPX for both methods). A custom, high NA objective (Special Optics 0.6 NA) is used to focus the tweezers into the vacuum chamber. Atomic fluorescence at the imaging wavelength is collected using the same objective and imaged onto an EMCCD camera (N\"{u}V\"{u} HN\"{u} 512 Gamma). For rearrangement, we use fast detection and feedback directly to the OPX using an intermediate module that processes the camera images into occupation matrices and communicates that information to the OPX in real time (Observe camera readout module, Quantum Machines).

\subsection*{Device Fabrication}
The nanophotonic chip is made from a 340 nm thick layer of commercially available stoichiometric LPCVD-grown silicon nitride (Si$_{3}$N$_{4}$) on top of a 600 $\mu$m thick silicon substrate (Silicon Valley Microelectronics). The nanophotonic devices are 1.1 $\mu$m wide and 63 $\mu$m  long and are repeated every 11 $\mu$m. The cavity design, electron-beam lithography (EBL), and reactive-ion etching (RIE) steps are carried out at the Center for Nanoscale Materials at Argonne National Laboratory, and the remaining processes and design are carried out in the Pritzker Nanofabrication Facility at the University of Chicago. The details of the fabrication are provided in the supplementary section.

\subsection*{Image Processing}
For the single-shot images labelled Image 1 and Image 2 in Fig. \ref{fig4}\textbf{b} and Fig. \ref{fig4}\textbf{c}, we first plot the two single-shot atom images and then plot a semitransparent image of the devices over them to show the device locations without obscuring the raw atom image data underneath. For Fig. \ref{fig3}\textbf{a}, Fig. \ref{fig4}\textbf{b} Image 2 Average, and Fig. \ref{fig4}\textbf{c} Image 2 Average, we take the average of the atom images and subtract from it an averaged background image (taken with identical imaging conditions after dropping the atoms from the tweezers). We then add a small offset value to ensure that all pixels in the background subtracted image are positive, if necessary, before multiplying the averaged, background-subtracted image by a scaling factor to match the values of the averaged atom image with the values of the device image. We then add this averaged atom image to a single image of the devices taken just prior to the experiment. This sum then forms the final plot shown.

\begin{acknowledgments}
We thank Kevin Singh, Yu-Hao Deng, Yuzhou Chai, Dahlia Ghoshal, Haley Nguyen, and Harry Levine for insightful discussions. We also thank Ramon Szmuk and Alex Kotikov for their help integrating the Observe and the OPX into our platform.

We gratefully acknowledge funding from the NSF QLCI for Hybrid Quantum Architectures and Networks (NSF award 2016136), the NSF Quantum Interconnects Challenge for Transformational Advances in Quantum Systems (NSF award 2138068), the NSF Career program (NSF award 2238860), and the Sloan Foundation. Photonic crystal cavity simulation, design, and device fabrication were performed at the Center for Nanoscale Materials, a U.S. Department of Energy Office of Science User Facility, supported by the U.S. DOE, Office of Basic Energy Sciences, under Contract No. DE-AC02-06CH11357.                                                                                            
\end{acknowledgments}

\bibliography{ref}

\clearpage

\onecolumngrid

\section*{Supplementary Information}

\renewcommand{\thefigure}{S\arabic{figure}}
\renewcommand{\figurename}{Fig.}
\setcounter{figure}{0} 

\begin{figure*}[b]
\includegraphics{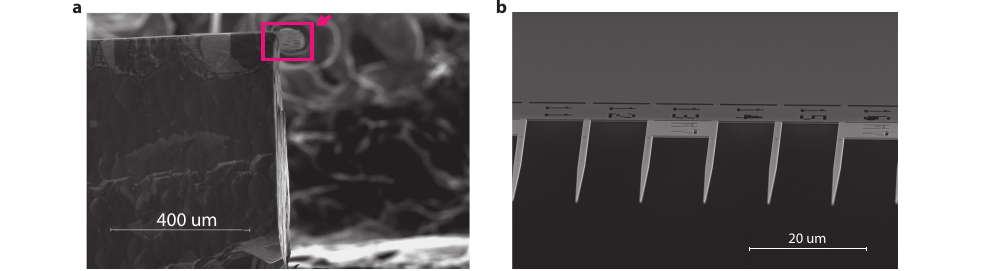}
\caption{\label{supplementary_fig_1} \textbf{SEM image of the chip with nanophotonic devices.} \textbf{a}, The devices are marked with the red box and are overhanging by 80 $\mu$m at the edge of the 600 $\mu$m thick chip. The devices are undercut through the entire chip thickness by etching the chip from the side. \textbf{b,} A zoomed in SEM view from the direction of the arrow in \textbf{a}.}
\end{figure*}

\section{Device Fabrication}
  The nanophotonic devices on the silicon nitride layer are patterned using conventional electron-beam lithography (EBL) with a ZEP 520A resist etch mask, followed by CHF$_{3}$/O$_{2}$ inductively coupled reactive-ion etching (RIE). Once the devices are patterned, the chip region of 2 mm $\times$ 8 mm around the devices is defined using photolithography with a AZ4620 resist etch mask. This is followed by deep reactive-ion etching (DRIE) and dicing to isolate the chip region. After protecting the silicon nitride device side using a double layer PMMA mask, a final KOH etching step is performed to undercut the silicon from the device region by etching from the sides. The undercut devices are shown in Fig. \ref{supplementary_fig_1}. 

\section{Loading Atoms onto Devices}
The atoms are loaded in the ``loading region" (as marked in Fig. \ref{fig2}\textbf{b}) and then translated along the X-direction until they are in between the devices. At this position, the Stark shift from the tweezer is similar to that of a free-space tweezer as depicted in Supplementary Fig. \ref{supplementary_fig_2}\textbf{a}. From here, atoms are loaded onto the devices by performing a Y-direction move, with a speed of 6 $\mu$m/ms. As the tweezer approaches the device, the partial reflection of the tweezer from the surface interferes with the incident beam causing a lattice to form, where each intensity maximum can trap atoms. The corresponding modified Stark shift on the atoms is shown in Fig. \ref{supplementary_fig_2}\textbf{b}. The proportion of atoms that load into various intensity maxima is optimized by adjusting the focal plane of the tweezer with respect to the device position. Previous work has shown up to 94\% loading efficiency into the closest intensity maximum of the standing wave potential, and Monte Carlo simulations show that up to 100\% loading efficiency can be achieved \cite{Thompson2013,Luan2020}. For the device thickness of 330 nm, used in this work, the maximum loading efficiency to the first intensity maximum from Monte Carlo simulations is $\sim$40\%. This loading efficiency is a function of the device thickness, atomic temperature, tweezer trap depth, angle of the devices with respect to the tweezer plane, and the adiabaticity of the tweezer movement while loading onto the devices. Further optimizations of these parameters can improve the loading efficiency. By performing Monte Carlo simulations of the Stark shifts experienced by the atoms in the various intensity maxima on top of the devices, we can estimate the proportion of atoms loaded into each of the intensity maxima. The dashed lines in Fig. \ref{supplementary_fig_2}\textbf{b} show the estimated Stark shifts experienced by atoms loaded into the three intensity maxima closest to the device surface (z1-z3). We fit our blowout curves probing the Stark shift on the devices assuming random sampling from these three curves to obtain the best-fit estimate of our loading distribution of 29\% to the intensity maximum closest to the device, 66\% to the second intensity maximum, and 5\% to the third.

\begin{figure*}
\includegraphics{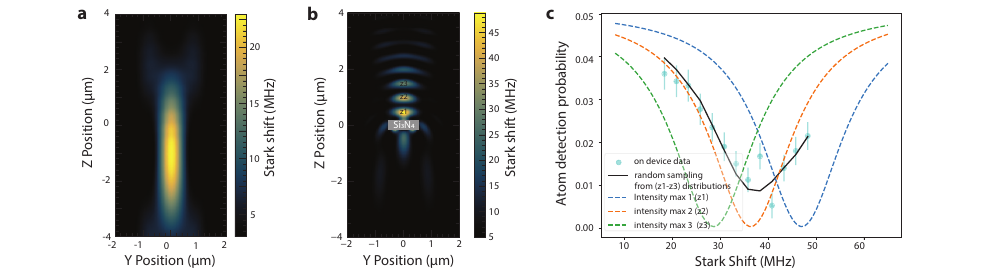}
\caption{\label{supplementary_fig_2} {\textbf{Stark shift.} \textbf{a}, Stark shift of the 895 nm D1 transition induced by a 2.4 mW, 935 nm optical tweezer with a waist of 1.1 $\mu$m. 
\textbf{b}, Modified trapping potential and the corresponding D1 Stark shift with the introduction of a silicon nitride device at the location marked. The tweezer is partially reflected by the device, resulting in the formation of a lattice-like potential with the closest intensity maximum (z1) causing the largest Stark shift.
\textbf{c}, Expected Stark shift survival curves for the first three intensity maxima (z1-z3). The experimental curve is fitted to random sampling from the three survival-curves, with a probability of 0.29 to be in the first intensity maximum, 0.66 to be in the second intensity maximum, and 0.05 to be in the third intensity maximum.}}

\end{figure*}

\subsection{Tweezer Objective Z Positioning and MOT Coil Stabilization}

The process of loading atoms onto the nanophotonic devices from free space is sensitive to the focal plane of the tweezers relative to the plane of the devices. Therefore, we require control over the Z position of the objective. Towards this end, we have our objective on a translation stage and use a piezoelectric screw to actuate the stage forwards and backward to vary the focal plane of the tweezers.

In order to optimize the position of the objective, we block the tweezers with a sliding optics mount holding a 900 nm LED which we turn on to illuminate the nanophotonic chip through our tweezer objective and image it on our EMCCD. We then assign a focal score to the image. We calculate the focal score by first applying a bilateral filter to the image to reduce the noise and then calculating the Laplacian of the filtered image to give us a numeric value that is proportional to the contrast in the image which is maximized when the image is in focus.

In general, we do not see the best loading results at the exact image focus, so we scan through the focus to establish the maximum score and target a focus score relative to the measured maximum in order to counteract long timescale drift in the LED output power. We can also choose a subsection of the image to target different planes due to the small relative angle between the chip and the objective.

During the course of the experiment, the variation in the focal score is monitored to ensure the stability of the Z-position reference. The focal position originally shifted by a few microns over the course of a couple of hours. The source of this variation was found to be the temperature gradient generated by the MOT coils during the experiment run time. To mitigate this problem, we stabilize the temperature around the coils and the objective region by maintaining a constant MOT on-off ratio during experiment cycles. When the experiment is not running the MOT coil currents are maintained to have the same effective heat output as running the experiment at the controlled duty cycle. With this careful temperature control in the vicinity of the objective and its translation stage, we achieve stability of the Z-position for up to 24 hours with less than 300 nm variation.

\section{Atomic Lifetime and Imaging On Devices}
\begin{figure*}[h]
\includegraphics{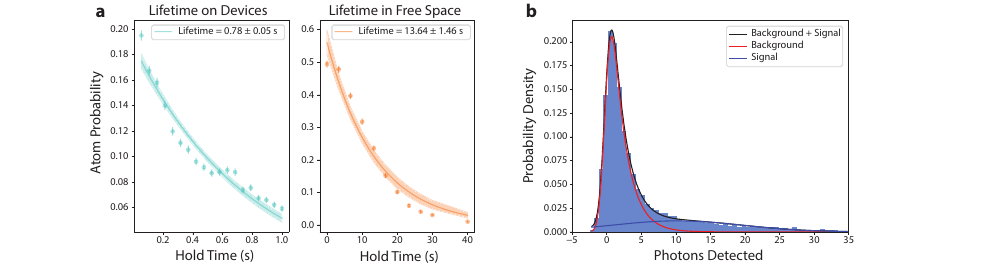}
\caption{\label{supplementary_fig_3} { \textbf{Atom lifetime and imaging on devices.} \textbf{a}, Atom survival in the tweezer as a function of holding time in the loading region (orange) and when trapped on top of the devices (cyan). Atoms are relatively long-lived even on top of the devices with lifetimes of individual atoms varying from 0.7 to 1 second. \textbf{b}, A typical single-atom imaging histogram while atoms are trapped on top of the devices. Here the fidelity of detection is estimated to be 86\% with a threshold of 5 photons. Optimization of the tweezer power and imaging parameters can further improve the imaging fidelity.}}
\end{figure*}

To characterize the atomic lifetime in the tweezers, we load atoms in the tweezers and hold them for a variable amount of time before imaging. The corresponding probability to detect the atom as the tweezer hold time is varied is shown in Fig. \ref{supplementary_fig_3}\textbf{a}. From the fits, we estimate an atomic lifetime of 13.6 seconds in the loading region and 0.78 seconds when trapped in the standing wave traps on top of the devices. Before the installation of the chip inside the chamber, atomic lifetimes in excess of 60 seconds in free-space optical tweezers were observed, indicating a finite reduction in the lifetime with the chip placement and the associated increase in the atomic flux required to create the MOT in the presence of the chip. 
While imaging fidelities in excess of 99\% are routinely obtained for atoms in the loading region, the imaging fidelity on the devices is reduced. An imaging histogram for an atom trapped on the device is shown in Fig. \ref{supplementary_fig_3}\textbf{c} demonstrating an imaging fidelity of 86\%. While the background is comparable to that of atoms in the loading region, the atomic signal is shifted to lower photon numbers. This is due to an increased atomic loss during the imaging due to the modified trapping potential. This can be further optimized in future experiments by optimizing the tweezer power as well as the imaging laser powers and detunings.

\begin{figure*}[b]
\includegraphics{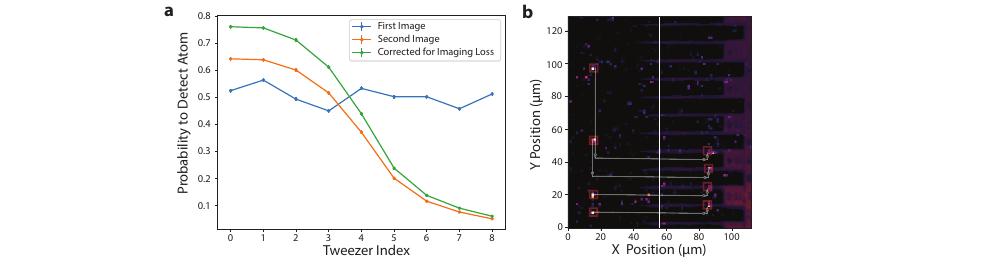}
\caption{\label{supplementary_fig_4} {\textbf{Rearrangement to the devices.} \textbf{a}, Rearrangement success in the compression part of the rearrangement to the devices. Stochastically loaded atoms (blue) are detected and compressed into a defect-free array (orange). Atom loss of 10-15\% during the first image results in a finite reduction of the atom probability after rearrangement. The green line shows the rearrangement curve after correcting for the imaging losses. \textbf{b}, The atomic trajectory used for rearrangement in Fig. \ref{fig4}. Atoms are first compressed into a defect-free array, then moved in between devices, and finally moved on top of the devices.}}
\end{figure*}

\section{Rearrangement}
Loading atoms into the optical tweezer array is a stochastic process with a 55\% loading probability. For deterministic placement of these atoms in a specific order onto a device or multiple devices, the stochastic loading pattern must be identified and moved to the target locations. We achieve this in a two-step process. In the first step, we identify the tweezers that were originally loaded with atoms and rearrange them into a defect-free configuration. In the second step, we move the defect-free array to the target location. In order to achieve a defect-free configuration of atoms, we first take an image of the stochastically loaded array to determine which tweezers have atoms and drop the tweezers that do not contain atoms. Following this, we chirp the remaining AOD frequencies that form our optical tweezers to compress the fully filled array into a defect-free configuration. We make this compression move in 1 ms following a symmetric piecewise quadratic frequency chirp profile. Blue points in Fig. \ref{supplementary_fig_4}\textbf{a} show the probability of stochastically loading each tweezer index. The orange points show the probability of detecting an atom at each tweezer site following the initial detection and rearrangement. A finite atomic loss during the first image and the losses due to the rearrangement result in non-unity atom probability after rearrangement. Around 10-15\% of atoms are lost after the first round of imaging, resulting in a corresponding reduction of atomic probability in the second image. The green curve in Fig. \ref{supplementary_fig_4}\textbf{a} shows the probability to detect atoms at each site after rearrangement, correcting for the site-specific atomic loss during imaging. The remaining 23\% atomic loss stems from the losses during rearrangement. The power imbalance between the tweezers during the process of rearrangement and specific frequency chirp profile can cause atoms to heat out of these traps during the first step. The imaging loss can be reduced by stroboscopic imaging and the rearrangement losses can be reduced by further characterization and optimization of the tweezer power, phase, chirp profiles, and trajectories enabling near unity rearrangement efficiencies.

After the first step of creating a defect-free array, atoms are then moved close to the target device using an X-direction move. This is followed by an adiabatic Y-direction move onto the devices. Fig. \ref{supplementary_fig_4}\textbf{b} shows the combined step 1 and step 2 trajectory of atoms. The left image shows the initial detection of the stochastically loaded atoms and their calculated trajectories. The right image shows the same atoms loaded on top of the device following the marked trajectories.

\end{document}